\documentstyle[prl,aps,multicol]{revtex}
\begin{document}
%\draft
\begin{multicols}{2}

\narrowtext

{\bf Schulz and Shastry Reply:}The previous Comment on our Letter Ref[1] (SS) solving exactly a class of two
chain fermi systems, draws attention to earlier work in Ref[2] where many 
lattice chains   are coupled together. The Comment makes several general
remarks, emphasizing the similarity of the resulting solutions, and expounds
on their significance in the light of  recent literature. We believe that the similar ``look '' of the
solutions notwithstanding, we are dealing with different classes of models.

The comparisions are facilitated by presenting the main points of our recent
generalization Ref[3] of the results of SS to the case of arbitrary number
of chains. Firstly consider the continuum: let $\alpha =1,\nu $ represent 
the index of one of $\nu $ chains of fermions $N_{\alpha }$ in number,
having coordinates $r_{i,\alpha }.$ The model solved is given by the
hamiltonian $H=\frac{1}{2}\sum_{\alpha ,i}\Pi _{i,\alpha }^{2}$, where the
canonical momentum $\Pi _{i,\alpha }=p_{i,\alpha }+\sum_{j,\beta }\xi
_{\alpha ,\beta }\,V_{\alpha ,\beta }(r_{i,\alpha }-r_{j,\beta })$, and the
object $\xi _{\alpha ,\beta }$ is antisymmetric $\xi _{\alpha ,\beta }=-\xi
_{\beta ,\alpha }=$sign$(\beta - \alpha  ).$ The gauge potential $V_{\alpha
,\beta }(x)=V_{\alpha ,\beta }(-x)=V_{\beta ,\alpha }(x)=$ $V_{\alpha ,\beta
}(x+L)$is obtained from  $E_{\alpha ,\beta }^{\prime }(x)=V_{\alpha ,\beta
}(x)$ satisfying $E_{\alpha ,\beta }^{{}}(x)=$ $E_{\beta ,\alpha }(x)=$ $%
-E_{\alpha ,\beta }^{{}}(-x).$ A pseudo unitary transformation maps H into a
``free model'': $\exp (iS\,)\,H\exp (-iS)=\widetilde{H}=\frac{1}{2}%
\sum_{\alpha ,i}p_{i,\alpha }^{2},$ where the generator $S=\frac{1}{2}%
\sum_{i,j,\alpha ,\beta }\xi _{\alpha ,\beta }\,E_{\alpha ,\beta
}(r_{i,\alpha }-r_{j,\beta }).$ Imposing PBC's on the wave function$\psi
=\exp (-iS)\psi _{free}(\{k_{i,\alpha }\}),$ gives us conditions on the
pseudo momenta $k_{i,\alpha }L=2\pi \,n_{i,\alpha }+\sum_{\beta }\xi
_{\alpha ,\beta }\,\Delta _{\alpha ,\beta }N_{\beta },$ where the phase
shift $\Delta _{\alpha ,\beta }=\int_{0}^{L}V_{\alpha ,\beta }(x)\,dx.$
Specalizing to $\nu =2$ recovers the prevous results in SS.

Next consider a collection of $\nu $ chains of fermions with a
lattice Hamiltonian $%
H_{l}=V-\sum_{m,\alpha }[(\exp i\phi _{m,\alpha })\,c_{m,\alpha
}^{\dagger}c_{m+1,\alpha }+h.c.]$, where $\phi _{m,\alpha }=\sum_{l,\beta }\xi
_{\alpha ,\beta }\,A_{\alpha ,\beta }(m-l)\,n_{l,\beta }.$  We perform a
unitary transformation with a suitably generalized $S=\frac{1}{2}%
\sum_{l,m,\alpha ,\beta }\xi _{\alpha ,\beta }B_{\alpha ,\beta
}(l-m)\,n_{l,\alpha }n_{m,\beta },$ with $B_{\alpha ,\beta }(m)=B_{\beta
,\alpha }(m)=-B_{\alpha ,\beta }(-m),$ where the burden of symmetry in
exchanging both spatial coordinates and species index is partitioned between 
$B_{\alpha ,\beta }$ and $\xi _{\alpha ,\beta }.$ We find $H_{l}^{\prime
}=\exp (iS)H_{l}\exp (-iS)=V-\sum_{\alpha }\{\sum_{m=1}^{L-1}[\,c_{m,\alpha
}^{\dagger}c_{m+1,\alpha }+h.c.]+[\exp i\Theta _{\alpha }\,c_{L,\alpha
}^{\dagger}c_{1,\alpha }+h.c.]\},$where the  twist angle $\Theta _{\alpha
}=\sum_{\beta }\xi _{\alpha ,\beta }\Delta _{\alpha ,\beta }\widehat{N}%
_{\beta }$,  the phase shift is $\Delta _{\alpha ,\beta
}=\sum_{m=1}^{L}A_{\alpha ,\beta }(m),$ and the number operator of fermions
in chain $\beta $ is $\hat{N}_{\beta },$ which can be replaced by its eigenvalue in
each sector, since it is conserved. Thus the model becomes a twisted b.c.
fermi chain with a given twist angle, and we can write down the equation
governing the pseudomomenta $Lk_{j,\alpha }=2\pi I_{j,\alpha }+\Theta
_{\alpha }+\{$Bethe Phase Shift$\}.$ In obtaining this simple form of $%
H_{l}^{\prime }$, we have chosen, as in SS the functions $A,B$ ($\alpha
,\beta $ in SS) to satisfy recursion relations $B_{\alpha ,\beta
}(m+1)-B_{\alpha ,\beta }(m)=A_{\alpha ,\beta }(m)$ for $1\leq m\leq L-1.$
This together with the oddness of $B_{\alpha ,\beta }$ forces a spatial 
symmetry on $A:$ $A_{\alpha ,\beta }(-m)=A_{\alpha ,\beta }(m-1).$ Once
again, restricting to $\nu =2$ recovers the results of SS.

{}From the above discussion of the lattice model, note that the {\em
minimal model}
of our class has $A_{\alpha ,\beta }=0$ except for $A_{\alpha ,\beta
}(0)=A_{\alpha ,\beta }(-1)\neq 0,$corresponding to the phase factor $\phi
_{m,\alpha }=\sum_{\beta }\xi _{\alpha ,\beta }\,A_{\alpha ,\beta
}(0)[\,n_{m,\beta }+n_{m+1,\beta }]$. The model of Ref[2] corresponds to $%
\phi _{m,\alpha }=\,$const$[\,n_{m,\alpha -1}-n_{m,\alpha +1}]$. While the
chain index dependence can be obtained by specializing our general $A,$ the
issue concerns the dependence on the spatial index. In order to obtain the
model of Ref[2], we must choose $A_{\alpha ,\beta }(m)=$const$\delta
_{m,0}\delta _{\alpha -\beta ,\pm 1}.$ This  spatial dependence is
inconsistent with the symmetry requirement noted in the previous paragraph,
and hence we find that {\it the model of }Ref[2] {\it cannot be solved
by our method. } Morover, there are important dfferences in the
boundary conditions in the chain direction, we have no need for any specific
bc's whereas the Ref[2] seems to. Therefore the dependence of the twist
angle $\Theta _{\alpha }$ on the number of particles in the boundary
chains are different from Ref[2].

To sumarize, the class of models introduced in SS contain considerable
freedom in regard to the range and nature of the interchain coupling,
subject however, to certain spatial-index
symmetries of the functions $A,B.$ The
models considered in Ref[2] do not satisfy our conditions, and we are
unable to solve the same. Detailed examination shows that the solutions
differ : {\it despite the  similar look, we are dealing with
different classes of models}Ref[4]. The details of the properties of the multi chain
models reported here will be published  in Ref[3]. BSS would like to thank
Bell Laboratories Lucent Technologies for hospitality,
where the 
work of Ref[3] was carried out.

\begin{flushleft}
H . J. Schulz$^1$ nd B Sriram Shastry$^2$\\
$^1$ LPS, University Paris-Sud, 91405 Orasy, France\\
$^2$ IISc, Bangalore 560012, India\\
\end{flushleft}

\end{multicols}

\begin{references}
\bibitem{1} H. J. Schulz and B. S. Shastry, Phys. Rev. Lett., 
{\bf 80}, 1924 (1998).
\bibitem{2} A. E. Borovik, A. A. Zvyagin, V. Yu. Popkov and Yu. M. 
Strzhemechnyi, JETP Lett., {\bf 55}, 292 (1992).;Also
 A. A. Zvyagin, Sov. J. Low Temp. Phys., {\bf 18}, 723 (1992). 
\bibitem{3} B. S. Shastry and H. J. Schulz, to be published (1998).
\bibitem{4} We use the term ``class of models'' in the specfic
sense that    the members share a common method of solution, rather than 
a vague sense in which many models share certain physical properties. 
\end{references}
\end{document}